\documentclass{pasa}%

\usepackage{graphicx}
\usepackage{amsmath}
\usepackage[utf8]{inputenc}
\usepackage[font=scriptsize]{caption}
\usepackage{subcaption}

\title{\Large \bf Estimating the values and variations of neutron star observables by dense nuclear
matter properties}
\author[P. P\'osfay et al.]{P\'eter P\'osfay$^{1}$, Gergely G\'abor Barnaf\"oldi$^1$ and Antal Jakov\'ac$^{1,2}$
\affil{$^1$Wigner Research Centre for Physics, P.O. Box, H-1525 Budapest, Hungary}%
\affil{$^2$Institute of Physics, E\"otv\"os University, 1/A P\'azm\'any P. Stny, H-1117 Budapest, Hungary}
}%

\jid{PASA}
\doi{10.1017/pas.\the\year}
\jyear{\the\year}

\usepackage{aas_macros}
\usepackage{hyperref} 
\hypersetup{colorlinks,citecolor=blue,linkcolor=blue,urlcolor=blue}
\usepackage{slashed} 
\usepackage{amssymb} 
\usepackage{hyperref}
\usepackage{xcolor} 
\usepackage{bm} 

\usepackage{booktabs}

\newcommand{\Lagr}{\mathcal{L}}

\newcommand{\mev}{\text{MeV}}
\renewcommand{\x}{\tilde{\sigma}}
\newcommand{\y}{\tilde{\omega}}
\newcommand{\z}{\tilde{\rho}}
\newcommand{\gs}{\tilde{g}_{\sigma}}
\newcommand{\gw}{\tilde{g}_{\omega}}
\newcommand{\gr}{\tilde{g}_{\rho}}
\newcommand{\dd}{\textrm{d}\,}

\begin{document}

\begin{frontmatter}
\maketitle

\begin{abstract}
Recent NICER observation data on PSR J0030+0451 has recently added a unique mass-radius constraint on the properties of the superdense nuclear matter existing in the interior of compact stars. Such a macroscopic data restrict further the microscopical models, elementary interactions, and their parameters, however, with reasonable margin because of the {\sl masquarade problem}. Here our goal is to identify the origin and quantify the magnitude of the theoretical uncertainties of the nuclear matter models.

A detailed study on the effect of different interaction terms and the nuclear parameter values in the Lagrangian of the extended $\sigma$-$\omega$ model is presented here. The equation of state was inserted to the Tolman\,--\,Oppenheimer\,--\,Volkoff equation and observable parameters of the neutron star were calculated. 

We identified, that the optimal Landau effective mass is the most relevant physical parameter modifying the macroscopic observable values. Moreover, the compressibility and symmetry energy terms just generate one-one order of magnitude smaller effect to this, respectively. We calculated the linear relations between the maximal mass of a compact star and these microscopic nuclear parameter values within the physical relevant parameters range. Based on the mass observational data we estimated the magnitude of the radii of PSR J1614$-$2230, PSR J0348+0432, and PSR J0740+6620 including theoretical uncertainties arising from the models' interaction terms and their parameter values choices. 
\end{abstract}

\end{frontmatter}
\section{Introduction}

The inner structure of compact astrophysical objects, such as magnetars, neutron stars, and perhaps quark or hybrid stars is an active multidisciplinary research area between astrophysics, gravitational wave physics, and nuclear physics. Recent NICER's measurement of the pulsar PSR J0030+0451 by~\cite{niser:2019} provided a new constraint on the properties of the superdense nuclear matter existing in the interior of compact stars. With such a precise observational data on mass and radius, nuclear matter modeling can be further restricted and one can achieve better consistency of the effective theories at microscopic level, applied widely to describe this dense region. Parameters corresponding to effective models are the interaction types, couplings, degrees of freedoms, and particle masses determine the macroscopic observables. However the opposite direction, i.e. deducing microscopical parameters form the macroscopic star properties, is not straightforward due to the {\sl masquarade problem}, and has relevant uncertainties. 

Beside NICER, further observatories: eROSITA, ATHENA, and SKA aim to measure the properties of compact stars~\cite{nasanicer,Ozel:2015ykl,merloni2012erosita,athena,ska}, and on the other hand, the discovered gravitational waves provide a channel to probe the inner structure of these celestial objects:~\cite{LIGO_NSNSGW_detection, Abott_gw1708_radius, Rezzolla:2016nxn}. These observations are especially important for the theoretical description of the dense nuclear matter, because: matter in this state can not be produced in particle accelerators, and first-principle calculations based on lattice field theory are still challenging at high chemical potentials present in compact stars~\cite{Katz_finite_mu_analitic_cont,katz_finite_mu_EoS,Katz_finite_mu_lattice}. Because of the above mentioned difficulties, effective theories still play an important role in studying the properties of cold dense nuclear matter~\cite{Holt:2014hma, Kojo:2017pfw}. Astrophysical measurements concerning compact stars can place constraints on these effective models and its parameter values in the region characterized by high density and low temperature~\cite{Ozel:2016,Raithel:2017ity}. 

Recent studies show the importance of the correct handling of the bosonic sector in effective theories of nuclear matter by~\cite{harmonic:2017,Zsolt:2007,Posfay:2017cor,KovacsP:2019,Kovacs:2021ger}. In this paper we study the connection between the parametrizations of effective nuclear models and measurable properties of compact stars. We also provide the theoretical uncertainties coming from the nuclear matter model types and their parameter values choice. Following the analysis in~\cite{Posfay:2017cor,symwal} we employ several modified versions of the $\sigma$-$\omega$ model in~\cite{norman1997compact,Schmitt:2010,Boguta0,Boguta1,Boguta2} which has three types of interaction terms in the bosonic sector. After calculating the equation of state (EoS) corresponding to different parametrizations of these models the mass-radius ($M$-$R$) diagrams are calculated by solving the Tolman\,--\,Oppenheimer\,--\,Volkoff (TOV) equations. Finally, we show how sensitive the $M$-$R$ diagram is to the differences in the bosonic sector and how the properties of compact stars (mass and radius) influenced by different saturation parameters of nuclear matter. 

\section{Extended $\sigma$-$\omega$ model in mean field approximation}

The extended $\sigma$-$\omega$ model describes the basic degrees of freedom which assumed to be present in neutron stars: protons, neutrons electrons in $\beta$-equilibrium. It also has enough -- but not too many --  parameters to facilitate the basic properties of the nuclear force by reproducing the saturation properties of the infinite nuclear matter. The Lagrange-function corresponding to the extended $\sigma$-$\omega$  model is
%
\begin{equation}
\begin{split}
\Lagr=
%
& \overline{\Psi} \left(
i \slashed{\partial} -m_{N} + g_{\sigma} \sigma -g_{\omega} \slashed{\omega}  + g_{\rho} \slashed{\rho}^{a} \tau_{a}
 \right) \Psi + \\
%
& +\frac{1}{2}\,\sigma \left(\partial^{2}-m_{\sigma}^2 \right) \sigma - U_{i}(\sigma) \\
%
&- \frac{1}{4}\,\omega_{\mu \nu} \omega^{\mu\nu}+\frac{1}{2}m_{\omega}^2 \, \omega^{\mu}\omega_{\mu} \\
%
&-\frac{1}{4} \rho_{\mu \nu}^{a} \, \rho^{\mu \nu \, a} + \frac{1}{2} m_{\rho}^2 \, \rho_{\mu}^{a} \, \rho^{\mu \, a} \\
%
&+ \overline{\Psi}_{e} \left(
i \slashed{\partial} - m_{e}
\right) \Psi_{e} \, ,
\end{split}
\label{eq:wal_lag}
\end{equation}
%
where $\Psi=(\Psi_{n},\Psi_{p})$ is the vector of proton and neutron fields, $m_{N}$, $m_{\sigma}$, and $m_{\omega}$ are the fermion, the sigma, and the omega meson masses and $g_{\sigma}$, $g_{\omega}$, and $g_{\rho}$ are the Yukawa couplings corresponding to the $\sigma$-nucleon,  $\omega$-nucleon, and $\rho$-nucleon interactions respectively. Furthermore,
%
\begin{equation}
\begin{split}
\omega_{\mu \nu} &=\partial_{\mu} \omega_{\nu}-\partial_{\nu} \omega_{\mu} \, , \\
\rho_{\mu \nu}^{a}& =\partial_{\mu} \rho_{\nu}^{a} - \partial_{\nu} \rho_{\mu}^{a} + g_{rho} \epsilon^{abc} \rho_{\mu}^{b} \rho_{\nu}^{c} \, 
\end{split}
\label{eq:rho-omega}
\end{equation}
%
are the kinetic terms corresponding to the $\omega$ and $\rho$ meson. Here $U_{i}(\sigma)$ 
is a general interaction term, which is different for the variants of the $\sigma$-$\omega$ model studied here for certain '$i$' values:
%
\begin{equation}
\begin{aligned}
U_{3}(\sigma) &= \lambda_{3} \sigma^{3} \, ,\\
U_{4}(\sigma) &= \lambda_{4} \sigma^{4} \, , \\
U_{34}(\sigma) &=\lambda_{3} \sigma^{3} + \lambda_{4} \sigma^{4}  \, .
\end{aligned}
\label{eq:U_types}
\end{equation}
%
The last term in eq.~\eqref{eq:wal_lag} corresponds to the free electrons, $\Psi_{e}$ which are needed to model the $\beta$-equilibrium and asymmetric nuclear matter present in neutron stars. 

The presence of the asymmetric nuclear components and electrons was not included in our previous calculation~\cite{Posfay:2017cor,symwal}. One of the goal of this extended work is to estimate the role of the asymmetry.

In the mean field (MF) approximation the kinetic terms are zero for the mesons and only the fermionic path integrals have to be calculated at finite chemical potential and temperature. The only components of the mesons which are not zero in the mean field approximation are $\omega_{0}=\omega$ and $\rho_{0}^{3}=\rho$. The calculation can be done as it is described in for example in~\cite{jakovac2015resummation}. The free energy corresponding to the model in the mean field approximation is given as: 
%
%
%
%
\begin{equation}
\begin{split}
f_{T} &=
%
 f_{F}
\left(
m_{N}-g_{\sigma} \sigma,
\mu_{p} - g_{\omega} \omega + g_{\rho} \rho
\right) \\
& +  f_{F}
\left(
m_{N}-g_{\sigma} \sigma,
\mu_{n} - g_{\omega} \omega - g_{\rho} \rho
\right)  \\
%
& + \frac{1}{2} m_{\sigma}^{2} \sigma^2 + U_{i}(\sigma) \\
%
& - \frac{1}{2} m_{\omega}^2 \omega^2 \\
%
& - \frac{1}{2} m_{\rho}^2 \rho^2 \\
%
& + f_{F} \left(m_{e}, \mu_{e} \right) \, ,
\end{split}
\label{eq:wal_f}
\end{equation}
%
where $\mu_{p}$, $\mu_{n}$, and $\mu_{e}$ are the proton, neutron, and electron chemical potential respectively. Note, muons are not included. The $f_{F}$ is the free energy contribution corresponding to one fermionic degree of freedom: 
%
\begin{equation}
\begin{split}
f_{F}(T,m,\mu) & = -2 T \int \frac{\dd^3 k}{(2 \pi)^3} 
\ln{\left[ 1 + \mathrm{e}^{-\beta \left( E_{k}-\mu \right) } \right]}  \,  , \\
\textrm{with} \ \ \ E_{k}^2 & = k^2 + m^2 \, ,
\end{split}
\end{equation}
%
which is in the model studied here describes the free energy of the proton and the neutron. In our calculations $T\to 0$ is used to describe the cold and dense nuclear matter of neutron stars. This means that the fermionic free energy has only two variables, $f_{F}(m, \mu)$.

During calculations it is useful to introduce the following notations for the expectation values of the fields:
%
\begin{equation}
\begin{split}
\x & = g_{\sigma} \sigma \, , \\
\y & = g_{\omega} \omega \, ,  \\
\z & = g_{\rho} \rho \,.
\end{split}
\end{equation}
%
These new variables in the expression of the free energy give a natural way to rescale the couplings:
%
\begin{equation}
\begin{split}
U_{i}(\sigma) & =  U_{i} \left( \frac{\x}{g_{\sigma}} \right) \, , \\
\gs & = \frac{m_{\sigma}}{g_{\sigma}} \, , \\
\gw & = \frac{m_{\omega}}{g_{\omega}}\, , \\
\gr & = \frac{m_{\rho}}{g_{\rho}} \, .
\end{split}
\end{equation}
%
With these new variables the free energy becomes the following:
%
\begin{equation}
\begin{split}
f_{T} &= f_{F} 
\left(
m_{N}- \x,
\mu_{p}- \y + \z
\right) \\
& + f_{F} 
\left(
m_{N}- \x,
\mu_{p}- \y - \z
\right) \\
& + \frac{1}{2} \gs^2 x^2 + U_{i}(\x) \\
& - \frac{1}{2} \gw^2 \y^2 \\
& - \frac{1}{2} \gr^2 \z^2 \\
& + f_{F} \left( m_{e}, \mu_{e} \right) \ .
\end{split}
\label{eq:wal_fxyz}
\end{equation}
%
The $f_{T}$ free energy depends on the electron, proton and neutron chemical potential and on the expectation values of the fields. The expectation values can be determined by the condition that in the physical point the free energy has extrema:
%
\begin{equation}
\begin{split}
%
0 = \frac{\partial f_{T}}{\partial \x}
= & \gs^2 \x + \frac{\partial U_{i}(\x)}{\partial \x} - \frac{\partial f_{F}}{\partial m} 
\left( m_{N}- \x, \mu_{p} - \y+ \z \right) \\
&-  \frac{\partial f_{F}}{\partial m} 
\left( m_{N}- \x, \mu_{n} - \y- \z \right) \\
%
0  = \frac{\partial f_{T}}{\partial \y}  
=& - \gw^2 \y - \frac{\partial f_{F}}{\partial \mu} 
\left( m_{N}- \x, \mu_{p} - \y + \z \right) \\
&- \frac{\partial f_{F}}{\partial \mu} 
\left( m_{N}- \x, \mu_{n} - \y - \z \right) \\
%
0  = \frac{\partial f_{T}}{\partial \z}  
=& - \gr^2 \z + \frac{\partial f_{F}}{\partial \mu} 
\left( m_{N}- \x, \mu_{p} - \y + \z \right) \\
&- \frac{\partial f_{F}}{\partial \mu} 
\left( m_{N}- \x, \mu_{n} - \y - \z \right) \ .
\end{split}
\label{eq:MF_eq}
\end{equation}
%
Here $-\frac{\partial }{\partial \mu} f_{F} =n_{F}(m, \mu)$ is the fermionic density and $\frac{\partial }{\partial m} f_{F}=n_{s}(m,\mu)$ is the so called scalar density.

\subsection{Solving the mean-field equations for symmetric nuclear matter}

The parameters of nuclear matter are given for the infinite symmetric nuclear matter in~\cite{norman1997compact}, hence to fit the model it has to be solved in the symmetric case first. Symmetry means that:
%
\begin{equation}
n_{p}=n_{n} \ .
\end{equation}  
%
This implies $\z=0$ and $\mu_{p}=\mu_{n}=\mu$ after substitution into equations~\eqref{eq:MF_eq}. There are no electrons present in the  nucleus which implies $n_{e}=0$ and $\mu_{e}=0$, which gives $f_{F}\left( m_{e}, \mu_{e} \right)=0$, so the free energy of the system~\eqref{eq:wal_fxyz} has no contributions from electrons. This way the these mean field equations are reduced to two equations and three variables: $\x$, $\y$, and $\mu$, the nucleon chemical potential. After specifying the value of the nucleon chemical potential, the equations can be solved numerically for the expectation values $\x$ and $\y$. Using the thermodynamic relations at zero temperature the energy density of the system can be calculated by:
%
\begin{equation}
\begin{split}
\epsilon &=- p + \mu n \, ,\\
f &=-p  \, ,
\end{split}
\end{equation}
%
where $\mu$ is the nucleon chemical potential and $n={n_{p} + n_{n}}={-2 \frac{\partial}{\partial \mu}f_{F}}$ is the nucleon density coming from the proton and neutron densities and $p = p_{n} + p_{p}$ is the nucleon pressure coming from the partial pressure values corresponding to each nucleon. 

\subsection{Solving the mean-field equations for asymmetric nuclear matter}
\label{sec:sym}

The equation system for the expectation values of the fields has six variables ($\x$, $\y$, $\z$, $\mu_{p}$, $\mu_{n}$, $\mu_{e}$) and only three equations. One additional equation comes from the condition that the star has to be electrically neutral, which implies: 
%
\begin{equation}
n_{p}=n_{e}
\end{equation}
%
Using the fact that $n_{F}=\frac{k_{F}^3}{3 \pi^2}$ where $k_{F}$ is the fermion's Fermi momenta and that $k_{F}^2=\mu_{F}^2 - m^2$, where $\mu_{F}$ is the fermion chemical potential and $m$ is the mass of the given fermion. The connection between the electron and proton chemical potential has the following form: 
%
\begin{equation}
\begin{split}
k_{e} &= k_{p} \\
\mu_{e}^2 - m_{e}^2 &= \left( \mu_{p} - \y + \z \right)^2 - \left( m_{N} - \x \right)^2 \, ,
\end{split}
\label{eq:NS_neutr}
\end{equation}
%
where $m_{e}$ and $\mu_{e}$ is the mass and the chemical potential of the electron, $\mu_{p}$ is the chemical potential corresponding to the proton and $m_{N}$ is the nucleon mass.
A further equation comes from the $\beta$-equilibrium: 
%
\begin{equation}
n\, \rightleftharpoons p \, + \, e \,.
\end{equation}
%
The above equation implies the following connection for the chemical potentials: 
%
\begin{equation}
\begin{split}
\mu_{n} = \mu_{p} + \mu_{e} \, .
\end{split}
\label{eq:asym2}
\end{equation}
%
The mean field equations~\eqref{eq:MF_eq} and the above equations together can be solved numerically as one equations system after the specification either of the three chemical potential values, because the other two are determined by the equations~\eqref{eq:NS_neutr} and~\eqref{eq:asym2}.

\section{Parameter fitting in the extended $\sigma$ - $\omega$ model}

The nuclear parameters are given for infinite symmetric nuclear matter, thus for the parameter fitting and determination of the couplings the mean-field equations are considered in the symmetric case. Here, no electrons are present and $ \z=0$ as it is described in Section~\ref{sec:sym}. All of the models considered here are fitted on the nucleon saturation data found in~\cite{norman1997compact,meng2016relativistic}. 
The definition of the Landau mass is, 
%
\begin{equation}
\begin{aligned}
m_{L} &=\frac{k_{F}}{v_{F}}  \quad \text{with} \quad
v_{F} &=\left.\frac{\partial E_{k}}{\partial k} \right|_{k=k_{F}} \, .
\end{aligned}
\label{eq:landau_mass}
\end{equation}
%
Where $k=k_{F}$ the Fermi-surface and $E_{k}$ is the dispersion relation of the nucleons. The Landau mass and the effective mass is not independent in relativistic mean field theories: 
%
\begin{equation}
\begin{split}
m_{L}= \sqrt{k_{F}^2 + m_{N, eff}^2} \, .
\end{split}
\label{eq:effmass_vs_landau_mass}
\end{equation}
%
This is the reason why the Landau mass and the effective mass of the nucleons can not be fitted simultaneously in the models we consider~\cite{meng2016relativistic}. In this paper we deal with this problem in the following way. We fit all of the models two times: 
\begin{enumerate}
\item[(i)] using the effective mass value from Table~\ref{tab:fitting_data} and
\item[(ii)] calculated from eq.~\eqref{eq:effmass_vs_landau_mass} to reproduce the Landau mass value from Table~\ref{tab:fitting_data}.
\end{enumerate}
%

%
\begin{table}[h]
\caption{\label{tab:fitting_data}Values of the nuclear parameters at the saturation point.}
\begin{center}
\begin{tabular}{ll}
\hline \hline 
\textbf{Parameter}              & \textbf{ Value}        \\
\hline
Binding energy $B$        & $-16.3$ MeV    \\
Saturation density, $n_{0}$    & $0.156$ $ \text{fm}^{-3}$  \\
Nucleon effective mass, $m^{*}$ & $0.6$  $m_{N}$      \\
Nucleon Landau mass $m_{L}$    & $0.83$ $m_{N}$ \\
Compressibility, $K$      & $240$ MeV   \\ 
Symmetry energy, $a_{sym}$      & $32.5$ MeV   \\   
\hline \hline 
\end{tabular}
\end{center}

\end{table}
%

The compression modulus or compressibility of nuclear matter is defined as,
\begin{equation}
\begin{split}
K =k_{F}^2 \frac{\partial^2 }{\partial k_{F}^2} \left( \frac{\epsilon}{n} \right)
= 9 n^2 \frac{\partial^2}{\partial n^2} \left( \frac{\epsilon}{n} \right) \, .
\end{split}
\label{eq:K}
\end{equation}
%
At saturation density the compression modulus has a simple connection to the thermodynamical compressibility at saturation density, $\chi$:
%
\begin{equation}
\begin{split}
\frac{1}{\chi}  = n \frac{\partial p}{\partial n} \quad \text{and} \quad
K = \frac{9}{n_{0}\chi} \, .
\end{split}
\end{equation}
%
The symmetry energy of nuclear matter is given as, 
%
\begin{equation}
a_{sym} = \frac{1}{2} \left. 
\frac{\partial^2 }{\partial t^2} \left( \frac{\epsilon}{n} \right) \right|_{t=0} \ \ ,
\label{eq:asym_def}
\end{equation}
%
where $t=\frac{n_{n}-n_{p}}{n_{B}}$. This value can be fitted only if the free energy~\eqref{eq:wal_fxyz} contains non-zero $\gr$ coupling. It's value does not effect the calculations when the nuclear matter is at saturation because in the infinite symmetric nuclear matter $\z=0$. It modifies the results when the equations are applied for neutron stars when the $\beta$-equilibrium forces proton--neutron asymmetry. 

If the models with $U_{3}(\sigma)$ and $U_{4}(\sigma)$ type interaction terms are used, than there is not enough free parameters to fit the data in Table~\ref{tab:fitting_data}. In these cases the nucleon effective mass, saturation density and binding energy are fitted and the compression modulus is the prediction. In the case of $U_{34}(\sigma)$ all four parameters can be fitted simultaneously and there is another way to incorporate data regarding both Landau mass and effective mass. For this model we consider a further fit where the value of the effective mass is chosen in a way that minimizes the error coming from not fitting the two types of masses correctly. Technically this value of the effective mass minimized the standard error of the fit. It's optimized value is:
%
\begin{equation}
\begin{split}
m_{opt}=0.6567 \, m_{N} \approx 616 \, \mev  \, .
\end{split}
\label{eq:optimal_mass}
\end{equation}
%
Since the compressibility is different for the three model variants with the given interaction terms they are compared in Table~\ref{tab:K_comp}. For models with $U_{3}(\sigma)$ and $U_{4}(\sigma)$ there are two fits, for Landau and effective mass which produce different compressibility values because they do not have enough free parameters to fit the correct value. However for $U_{34}(\sigma)$ there are enough parameters to fit the compressibility so it has the same value for all three fits: for the Landau mass, for the effective mass and for the optimal mass. These three parametrizations of the same model differ in their predictions for higher densities of nuclear matter.

Although these compressibility values are higher than the best estimates from Skyrme-potetnial based measurements in Refs.~\cite{Colo:2004,Klahn:2006}, consequently such conditions can help us to choose the most physically-reliable model, $U_{34}(\sigma)$. Nevertheless, all considered models result in similar descriptions at potential level. 
%
\begin{table}[h!]
\caption{\label{tab:K_comp} The value of the compressibility corresponding to different interactions in the extended $\sigma$ - $\omega$ model defined by the Lagrangian~\eqref{eq:wal_lag} and scalar interaction terms~\eqref{eq:U_types}. All model types are parametrized to reproduce either the Landau mass or nucleon effective mass values given in Table~\ref{tab:fitting_data} because these nuclear parameters can not be fitted simultaneously.}
\begin{center}
\begin{tabular}{ll}
\hline \hline 
\textbf{Interaction term}              & \textbf{$K$ [MeV]}        \\
\hline
original $\sigma$ - $\omega$ model & 563    \\
$U_{3}(\sigma)$ with effective mass fit    & 437   \\
$U_{3}(\sigma)$ with Landau mass fit &  247  \\
$U_{4}(\sigma)$ with effective mass fit  & 482 \\
$U_{4}(\sigma)$ with Landau mass fit     & 334   \\  
$U_{34}(\sigma)$ with Landau and effective mass   & 240   \\ 
\hline \hline  
\end{tabular}
\end{center}

\end{table}
%
\subsection{Determination of the couplings}

To fit the couplings first the expectation values of the fields has to be determined at the saturation point. The $\x_{0}$ is the expectation value of the rescaled sigma meson field and it can be determined after the specification of the effective nucleon mass at saturation, using the defining equation of the effective nucleon mass:
%
\begin{equation}
m_{N,eff}=m_{N}- \x_{0} \, .
\end{equation}
%
The $\y_{0}$ expectation values of the rescaled $\omega$ field at saturation can be determined from the condition that the binding energy $B=\frac{\epsilon}{n}-m_{N}$ must have a minimum at saturation density: 
%
\begin{equation}
\begin{split}
0&=\frac{\partial B}{\partial n}=\frac{\mu n- \epsilon}{n^2} \\
\mu_{0} &= B + m_{N} \, .
\end{split}
\end{equation}
%
This has to be substituted into to the definition of the effective chemical potential:
%
\begin{equation}
\begin{split}
\mu_{0}^{*} = \mu_{0} - \y_{0} 
= \sqrt{m_{0}^{* \, 2} + k_{F, 0}^2}  \ \ ,
\end{split}
\end{equation}
%
where $k_{F,0}$ is the nucleon Fermi momenta at the saturation density, which can be obtained from the nuclear density at saturation: $n_{0}=\frac{2 k_{F,0}^3}{3 \pi^2}$

The $\gw$ coupling can be determined from the second equation in equation system \eqref{eq:MF_eq} taking into account that $\z=0$ and $-2\frac{\partial }{\partial  \mu} f_{F} =n_{0}$ obtaining, 
%
\begin{equation}
\gw^2=\frac{n_{0}}{\y_{0}} \, .
\end{equation}
%
The self couplings of the $\sigma$-meson can be calculated from the first equation in the equation system~\eqref{eq:MF_eq} by setting $\z=0$ and $n_{s,0}=n_{s}(m_{0}^{*},\mu_{0})$: 
%
\begin{equation}
2 n_{s,0} = \gs^2 \x + \frac{\partial U_{i}(\x) }{\partial \x}   \ . 
\label{eq:dp/dx=0} 
\end{equation}
%
If $U_{i}(\sigma)=U_{3}(\sigma) \, \text{or} \, U_{4}(\sigma)$ in the Lagrangian \eqref{eq:wal_lag} this equation contains two variables. A further equation is provided by the condition that ${p=-f=0}$ at saturation. Making the necessary substitutions in equation~\eqref{eq:wal_fxyz} the second equation required for the couplings is obtained with $\z=0$ and $f_{F}\left( m_{e},\mu_{e} \right)= 0 $, 
%
\begin{equation}
\begin{split}
0 &= 2 f(m_{0}^{*},\mu_{0}-\y_{0})  
+ \frac{1}{2} \gs^2 \x_{0}^2 + U_{i} \left( \x_{0} \right) 
- \frac{1}{2} \gw^2 \y_{0}^2 \ .
\end{split}
\label{eq:p=0}
\end{equation} 
%
Equations~\eqref{eq:dp/dx=0} and~\eqref{eq:p=0} has to be solved together to determine the remaining couplings. If $U_{i}(\sigma)=U_{34}(\sigma)$ there are three couplings to be determined. In this case after solving~\eqref{eq:dp/dx=0} and~\eqref{eq:p=0}  one coupling is still undetermined. This freedom can be used to fit the compression modulus of the nuclear matter at saturation density. 

The coupling $\gr$ can be  determined as it is described in~\cite{norman1997compact} using the definition of symmetry energy in eq.~\eqref{eq:asym_def} : 
%
\begin{equation}
\begin{split}
a_{sym} = \frac{k_{F}^3}{3 \pi^2}
\frac{1}{\gr ^2} 
+ \frac{k^2_{F}}{6 \sqrt{k_{F}^2 + m^{* \, 2}}} \, .
\end{split}
\end{equation}
\section{Nuclear equation of state for asymmetric matter}

In this section equation of states corresponding to different parametrizations of the asymmetric extended $\sigma$-$\omega$ model given by eq.~\eqref{eq:wal_lag} are compared. As mentioned above parameters describing nuclear matter are fitted with the appropriate symmetric nuclear matter data. These parameters then are used  when the effect of $\beta$-equilibrium is taken into account. The equation of state and nuclear properties corresponding to the symmetric version of the same models and parametrizations studied here can be found in~\cite{symwal}.

Models with three types of interaction terms ($U_{3}(\sigma)$, $U_{4}(\sigma)$, and $U_{34}(\sigma)$) are considered using two parametrizations corresponding to Landau mass and effective mass for each variant. In the case of the model characterized by the  interaction term $U_{34}(\sigma)$, a third type of fit was applied in addition using the effective mass value from eq.~\eqref{eq:optimal_mass}. 

\begin{figure}[!h]
\centering
\includegraphics[width=0.50\textwidth]{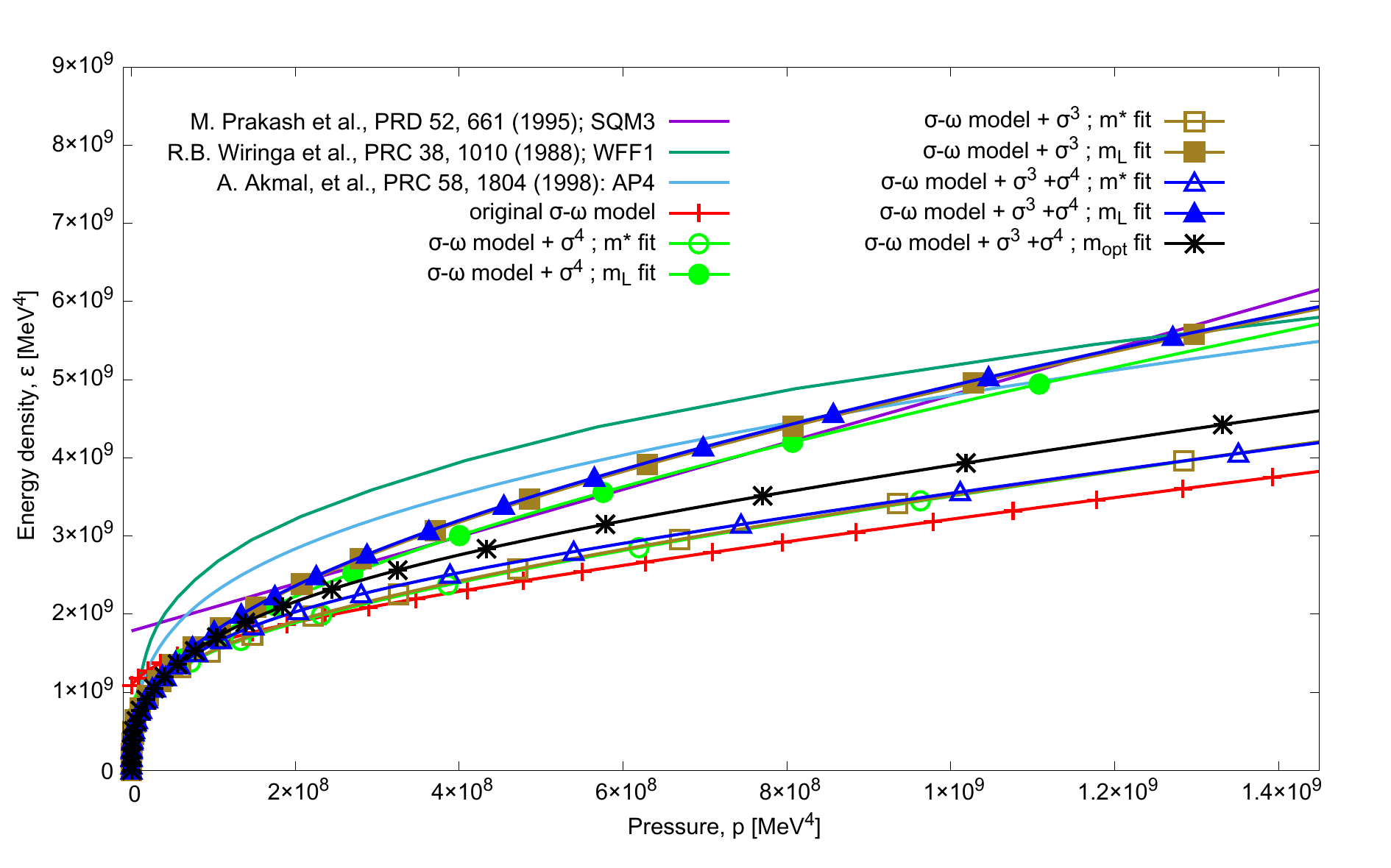}
\caption{\label{fig:eos_comp}
The EoS of the different parametrizations of the $\sigma$-$\omega$ model for asymmetric matter. The EoS parametrized to reproduce Landau mass has filled symbols, and the other EoS which are fitted on the effective mass has open symbols. For comparison the SQM3, AP4 and WFF1 equation of state was taken by~\cite{sqm3,ap4,wff1}. 
}
\end{figure}

The energy density, pressure, and density is calculated in all of these models. The calculated equation of states corresponding to these models are shown on Fig.~\ref{fig:eos_comp}. The results from the extended $\sigma$-$\omega$ model are compared to  the SQM3, AP4 and WFF1 equaiton of state taken from~\cite{sqm3,ap4,wff1}. For the sake of completeness and comparability the EoS of the original $\sigma$-$\omega$ model is also show on Fig.~\ref{fig:eos_comp}. 
%
%

\newpage
There are several important features of the EoS diagrams show on Fig.~\ref{fig:eos_comp}, listed here:
\begin{itemize}
\item The different EoS' are arranged on the $p$-$\epsilon$ plane according to the value of the effective mass. Regardless of the interaction type used in the extended $\sigma$-$\omega$ model the EoS of the models fitted on the same value of the effective nucleon mass, $m_{N, eff}$ run close to each other. According to this, the EoS curves presented on Fig.~\ref{fig:eos_comp}, form two bands corresponding to the possible ways to fit the models on the data presented in Table~\ref{tab:fitting_data}. The models are either parametrized to reproduce the value of the Landau mass $m_{L}$ or the nucleon effective mass $m^*$. The EoS of the model which is fitted on the optimal nucleon mass $m_{opt}$ runs between the two bands of EoS mentioned above, corresponding to the fact that the optimal nucleon mass is between the effective and Landau mass. 
\item The above mentioned behaviour of the EoS is present in the symmetric ${\sigma \text{-} \omega}$ model as well \cite{symwal}. This means that the inclusion of the $\beta$-equilibrium affect the different model variants in a similar way and diagrams retain their relative positions to each other. 
\item There is weak correspondence between the value of the compression modulus and the behaviour of the EoS on Fig.~\ref{fig:eos_comp}. This can be seen clearly by comparing the EoS of the models with the $U_{34}(\sigma)$ type interaction term with the  value of the compression modulus in the models which can be found in Table~\ref{tab:K_comp}. The value of the compression modulus in all cases is $240 \, \text{MeV}$ and the only difference between the models is the value of the effective nucleon mass, $m_{N,eff}$. The graphs of the EoS are arranged according to the values of the effective nucleon mass and they are far apart on the ${p \text{-} \epsilon}$  plane despite the fact that the value of the compression modulus is the same. 
\end{itemize}

To summarize, the value of the effective mass influences the behaviour of the EoS more than the compression modulus and the symmetry energy. If the model is parametrized to reproduce the Landau mass then at given pressure the energy density is higher than it would be in the same model if it was parametrized to reproduce the nucleon effective mass value $m^*$. The EoS of different variants of the $\sigma$-$\omega$ model with different interaction terms are very close to each other if the value of the nucleon effective mass is the same in the models, despite the fact that the compression modulus has different values in these model variants. 

\section{Solving the Tolman\,--\,Oppenheimer\,--\,Volkoff equations}

The Tolman\,--\,Oppenheimer\,--\,Volkoff equations (TOV) are used to calculate the properties of neutron stars corresponding to the models presented on Fig.~\ref{fig:eos_comp}. The TOV equations give the general relativistic description of the compact stars assuming spherically symmetric space time and time independence as it is described in~\cite{norman1997compact,Haensel_book}: 
%
\begin{equation}
\begin{aligned}
& \frac{\mathrm{d} p(r)}{\mathrm{d} r} =- \frac{G \epsilon(r) m(r)}{r^2} \times \\
& \qquad \left[ 
1 + \frac{ p(r)}{\epsilon(r)}
\right]
\left[
1 + \frac{4 \pi r ^3 p(r)}{m(r)}
\right] 
 \left[
1 - \frac{2 G m(r)}{r}
\right]^{-1} \\[10pt]
 & \frac{\mathrm{d} m(r)}{ \mathrm{d} r } =4 \pi r^2 \epsilon(r)
\end{aligned}
\label{eq:TOV}
\end{equation}
%
where $p(r)$  and $\epsilon(r)$ are the pressure and energy density as functions of the radius of the star, $G$ is the gravitational constant while $m(r)$ is the mass of the star which is included in the mass shells up to the radius, $r$. To integrate the equations one need a connection between $p(r)$ and $\epsilon(r)$ at given $r$, which is provided by the EoS in the form of the relation $p(r)=p(\epsilon(r))$. To start the integration one has to choose a central energy density values, $\epsilon_{c}$ for the star as an initial condition. 

It is unrealistic to use the EoS presented above, during the full integration of the TOV equations, because the crust of the neutron stars is not described by the nuclear equation of state by~\cite{Haensel_book,Chamel_crusts}. This problem can be solved by switching to a crust equation of state at the lower pressures where the nuclear EoS is no longer valid~\cite{norman1997compact,Haensel_book}. We note that any kind of inclusion of the crust equation of state introduces a new variable in the analysis and makes it more challenging to study the effect of nuclear equation of state on neutron star observables. To circumvent this in our analysis we used a different method.

\subsection{The core approximation}
To calculate the $M$-$R$ diagram of neutron stars in this paper a conservative approximation is employed. The integration of the TOV equations is stopped when the crust of the neutron stars is reached. This means that the part corresponding to the crust is never calculated thus the mass and the radius values resulting from the calculations is smaller than it would be if the integration was continued. The advantage of this method is that it allows a reasonable approximation of the $M$-$R$ diagram while it does not introduce the effect of the crust equation of state and nuclear equation of state is used only where it is valid. These features are important in this analysis as the aim of this paper is to study the relations between microscopic parameters of nuclear matter and neutron star observable like mass and radius. The inclusion of crust equation of state introduces other parameters which are not belong to the nuclear EoS describing the core of the star. For the sake of simplicity the calculation method described above is referred as {\sl core approximation} in this paper. We are aware that crust can play a crucial role in determining physical properties of small mass neutron start, $M<M_{\odot}$. Our strategy is to consider larger-mass compact objects, where this effect is diminished~\cite{lotmodell}.

\subsection{Remarks on the crust}

There is still one parameter in the core approximation which is introduced by the crust EoS. This is the value of the pressure called $p_{0}$ where the integration of the TOV equations is stopped. In the calculations presented in this paper the BPS equation of state is used to determine the pressure value where the core of the neutron star ends and the integration of the TOV equations is stopped~\cite{BPS}. The results presented here are representative as it is found that in the core approximation the precise value of the pressure where the integration is stopped is not that important, from the point of view of the $M$-$R$ diagram. The BPS equation of state only serves as a guide to set the correct order of magnitude for $p_{0}$. 

\begin{figure}[!h]
\centering
\includegraphics[width=0.49\textwidth]{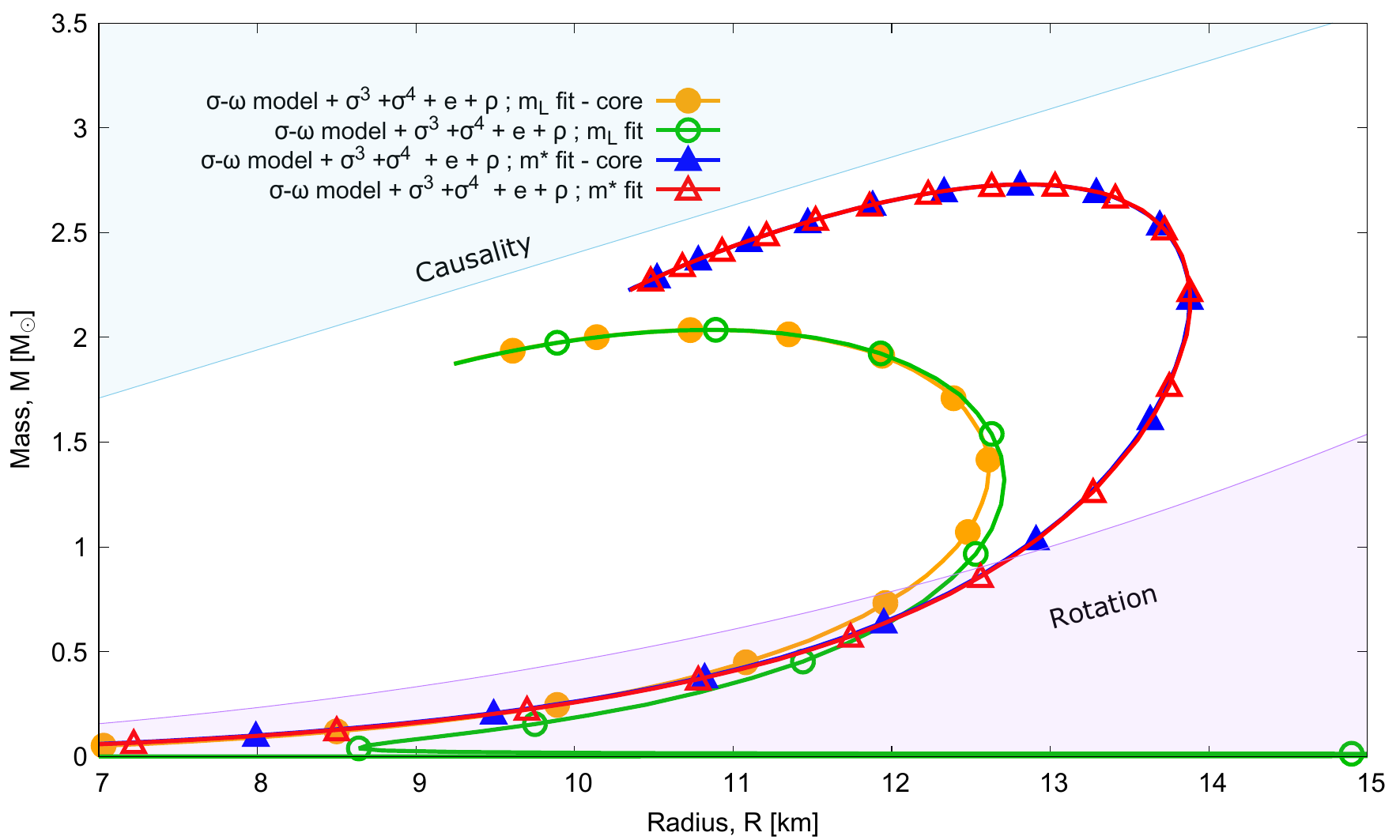}
\caption{
\label{fig:coreappr}
$M$-$R$ diagrams corresponding to the extended $\sigma \text{-} \omega$ model calculated with two different methods. The circle and triangle marker corresponds to the model variants fitted on the $m_{L}$ Landau mass and on the $m^*$ effective nucleon mass. The graphs with filled symbols correspond to core approximation and the hollow ones show the result of the full integration of the TOV equations using only the core nuclear equation of state.  
}
\end{figure}

The above mentioned core approximation is a conservative method to calculate the $M$-$R$ diagram, because the crust is not taken into account. To better understand the difference between the core approximation and the original method, where integration of the TOV equations is continued through the neutron star crust, the results of the two methods is compared on Fig.~\ref{fig:coreappr}. Two types of $M$-$R$ diagram are used in the comparison. Both of them predict that as the mass of the compact object become smaller the radius is also decreasing, but as the star mass becomes really small, $\lesssim 0.3 \, M_{\odot}$, one type of curve continues this trend, and the other turns back, and predicts larger and larger radius as the mass of the neutron star are becoming smaller and smaller. These two types of $M$-$R$ diagram behaves differently in the core approximation. In the case of the diagram which continues the original trend the core approximation and the original method has smaller difference than the line width on the diagram. The $M$-$R$ diagram which turns back from the original trend, and predicts larger radii at smaller masses, behaves differently: the difference between the two calculations is around 1-3\% for middle and low mass neutron stars and grows as the star mass becomes smaller.  At high mass stars however the difference is smaller than the line width. 

To summarize, the calculation based on the core approximation is very precise prediction for the high-mass neutron stars and in the worst case scenario it predicts a few percent smaller radius for mid- and low mass neutron stars. 
%

\section{Compact stars in the effective models}

\begin{figure}[!h]
\begin{center}
\includegraphics[width=0.49\textwidth]{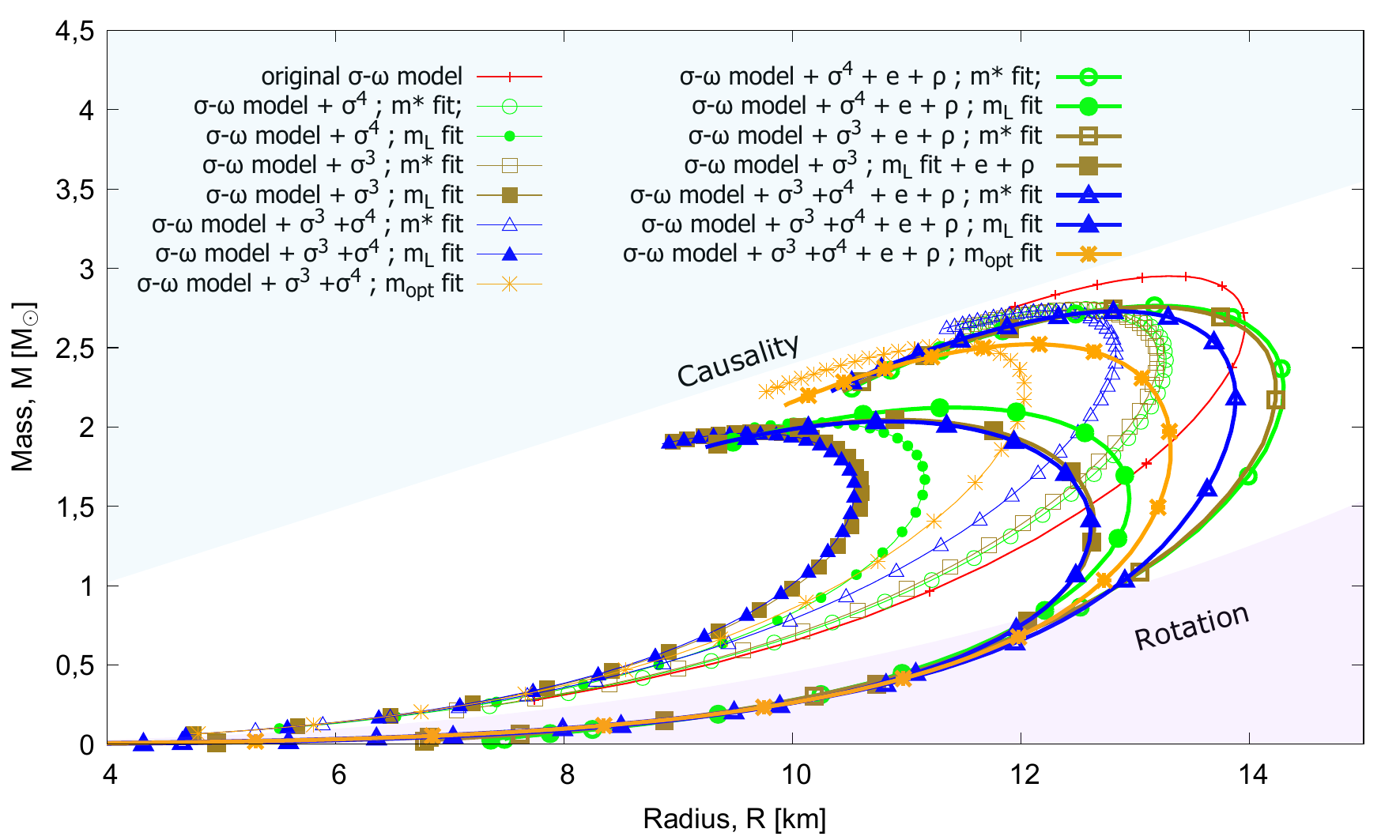}
\end{center}
\caption{\label{fig:mr}
Mass-radius diagram for the different parametrizations of the modified $\sigma$-$\omega$ model. The line representing the original $\sigma$-$\omega$ model is also drawn in red for comparison. Dashed color lines correspond to the assumption of symmetric nuclear matter and are taken from~\cite{symwal} and solid color lines correspond to matter in $\beta$-equilibrium. The same models have the same type of symbols and colors on their respective curves.}
\end{figure}

After solving the TOV equations using the method described above the mass and radius of a compact star with a given energy density can be determined. Results corresponding to different energy densities in a given model are summarized on a mass-radius diagram. We have calculated the $M$-$R$ diagram assuming nuclear matter in $\beta$-equilibrium in neutron stars and compared our results to calculations assuming symmetric infinite nuclear matter~\cite{symwal}. In the symmetric case we consider the same models without the $\beta$-equilibrium which permits us to directly compare the symmetric and asymmetric cases and study the effect of assuming asymmetric matter on the $M$-$R$ diagram. The results are shown on Fig.~\ref{fig:mr}. 
%
%
The model variants in both symmetric and asymmetric case show similar behaviour as the EoS on Fig.~\ref{fig:eos_comp}: the curves separate based on whether or not they are parametrized to reproduce the effective mass or Landau mass. Models with smaller effective mass systematically produce higher maximum mass stars compared to their parametrization with larger effective mass (Landau mass). Moreover all models fitted for the effective mass value in Table~\ref{tab:fitting_data} produce higher maximum star mass than the ones fitted for the Landau mass. Since the Landau mass and effective mass are not independent~\eqref{eq:landau_mass} the above statement is equivalent to saying that higher effective mass produces smaller maximum star mass. This picture is supported by the curve corresponding to the model variant parametrized by the optimal mass eq.~\eqref{eq:optimal_mass}, which is the best fit of the model. The maximum star mass in this case is between the values produced by parametrizations described by smaller and larger effective nucleon mass than the optimal mass value.

The above means that including $\beta$-equilibrium and asymmetry in the extended $\sigma$-$\omega$ model changes the $M$-$R$ diagram in a consistent way ({\sl solid color lines}). The relative positions of the curves remains as it was in the symmetric case ({\sl dashed color lines}). They retain the separation of the $M$-$R$ curves corresponding to the effective mass as discussed above. If a model in the symmetric case produces higher maximum star mass than another than this will be the same for the asymmetric case. A similar observations can be made for the maximum radius of the stars. 

The effect of including $\beta$-equilibrium on a single $M$-$R$ also clear: it increases the maximum neutron star mass and the radius compared to pure symmetric nuclear matter case. The effect of including asymmetry is much larger on the neutron star radius than on the maximum neutron star mass. This means that very precise mass and radius measurements are needed to detect effect of asymmetric nuclear matter on the $M$-$R$ diagram.  

It is important to note that parametrizations corresponding to the effective mass and to the optimal mass values are ruled out by observations as the produce too high maximum star mass and radius in both symmetric and asymmetric cases~\cite{Ozel:2016oaf}. The parametrizations corresponding to the Landau-mass are also very close to the limiting maximal neutron star mass. This means that in the case of the extended $\sigma $-$\omega $ model higher values of Landau mass (effective mass) are needed to reproduce the neutron star measurements  than nuclear physics data in Table~\ref{tab:fitting_data} would suggest. This can be reconciled by the fact that the extended $\sigma$-$\omega$ model is effective and it's parameters are not strictly translate into the measured properties of nuclear matter. It is also worth to mentions, that it is known that including higher mass nucleons, hyperons in the EoS reduces the maximum neutron star mass~\cite{norman1997compact}. It is possible that including heavier hadrons in the $\sigma$-$\omega$ model will balance the effect of effective nucleon mass and would produce neutron stars in the correct maximum mass range using smaller Landau mass values. 
 
The value of compressibility in these models seems to influence very little the maximum star mass, but at the same time it has an effect on the compactness by influencing the radius of the star which can be seen on Fig.~\ref{fig:mr} by looking at the curves belonging to the fits with the same effective mass (Landau mass). In these models all of the parameters are the same which listed in Table~\ref{tab:fitting_data} apart from the compressibility. 

\section{Connection between compact star observables and nuclear parameters}

So far we have presented the strong connection between the nuclear effective mass and maximum star mass. The model variant with interaction term $U_{34}(\sigma)$ is chosen to further investigate this phenomena and to study how the symmetry energy and the compression modulus influence the $M$-$R$ diagram. This model variant has enough free parameters to fit all of the nuclear saturation data listed in Table~\ref{tab:fitting_data} except the Landau mass and effective mass which can not be fitted simultaneously. To determine how the values of the microscopic parameters of nuclear matter influence the $M$-$R$ diagram various models fits are calculated with different values of nuclear parameters. 

The calculations are done in a following way. First the nuclear parameter is chosen which will vary while the others will be kept constant. For every value of the changing nuclear parameter the model couplings are recalculated so it reproduces the other constant parameters which does not change. This way for every value of the changing parameter there is a (slightly) different EoS. The $M$-$R$ diagrams corresponding to these EoS are calculated solving the TOV-equations. 

We have picked the most characteristic point of a typical $M$-$R$ diagram to study how neutron stars change as the nuclear parameters change.  As it  can be seen on Fig.~\ref{fig:mr} on every $M$-$R$ curve exits a star which mass is the largest, the Maximum Mass Star (MMS). The mass and radius of a MMS is determined as functions of the nuclear parameters and the $M$-$R$ diagrams are calculated from the EoS associated to the different values of the nuclear parameters. 

\begin{figure*}[!h]
\begin{subfigure}{0.5 \textwidth}
\centering
\includegraphics[width=0.95\linewidth]{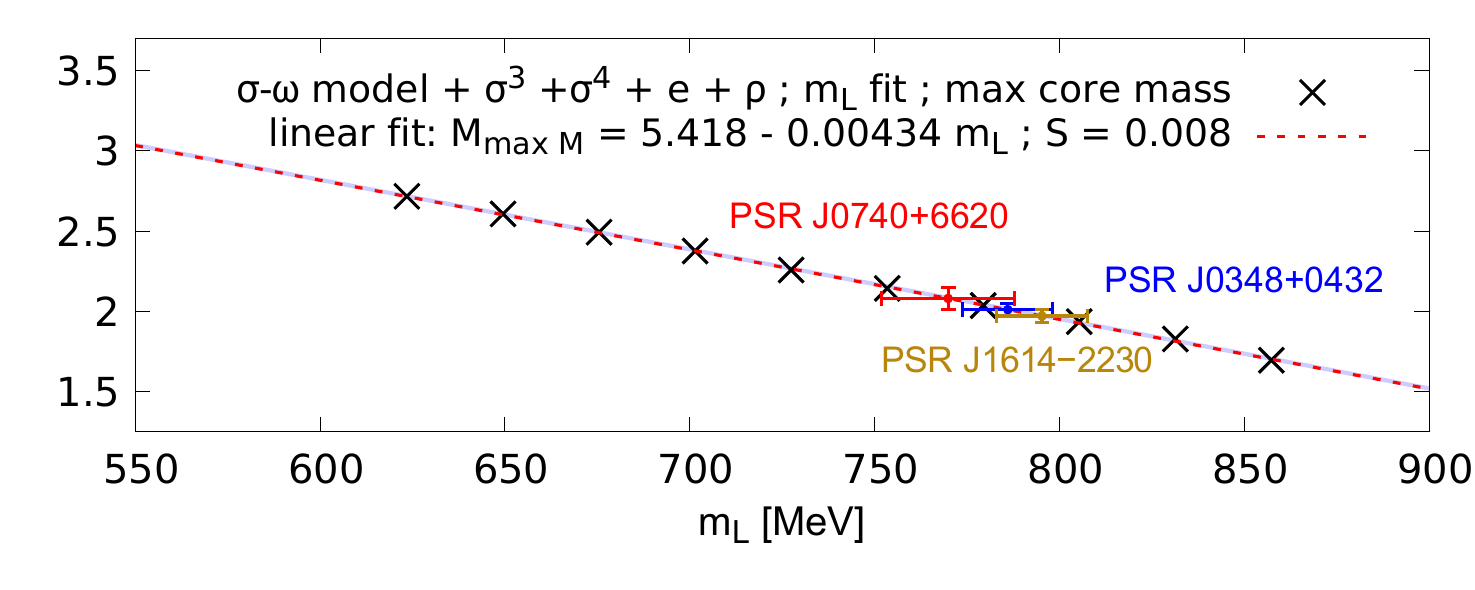}
\caption{}
\label{sfig:mm_mL_m}
\end{subfigure}
\begin{subfigure}{0.5 \textwidth}
\centering
\includegraphics[width=0.95\linewidth]{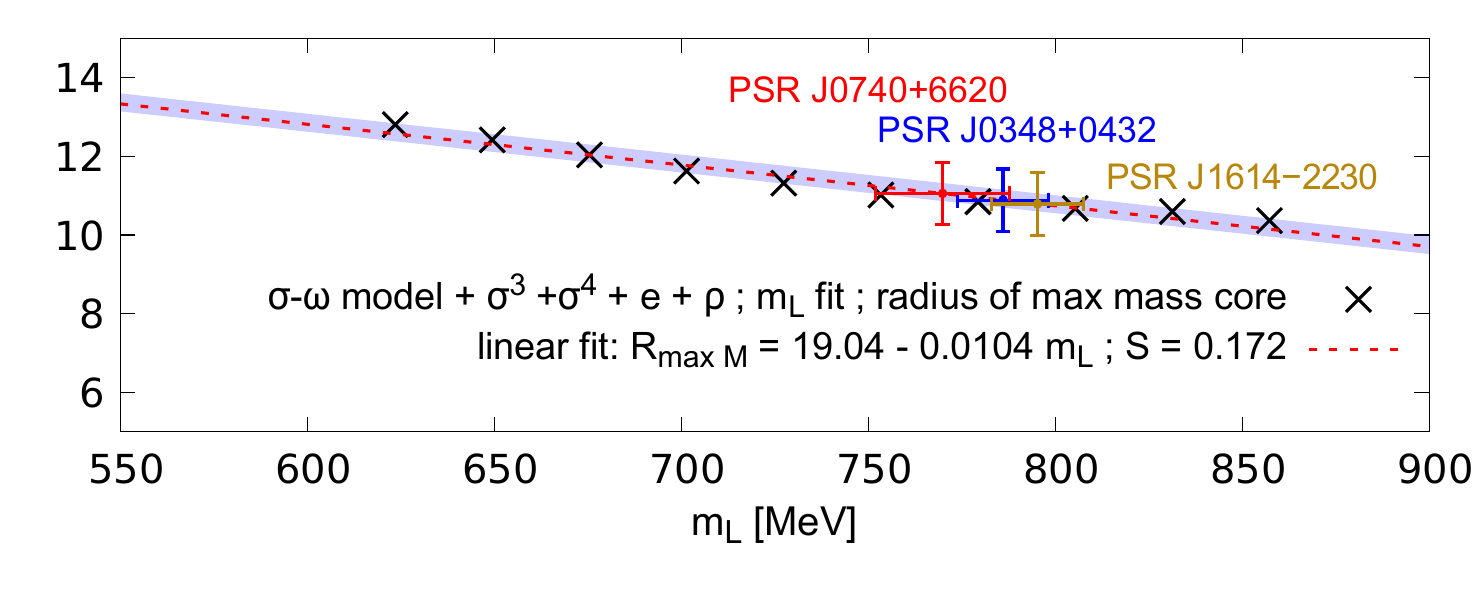}
\caption{}
\label{sfig:mm_mL_r}
\end{subfigure}
\begin{subfigure}{0.5 \textwidth}
\centering
\includegraphics[width=0.95\linewidth]{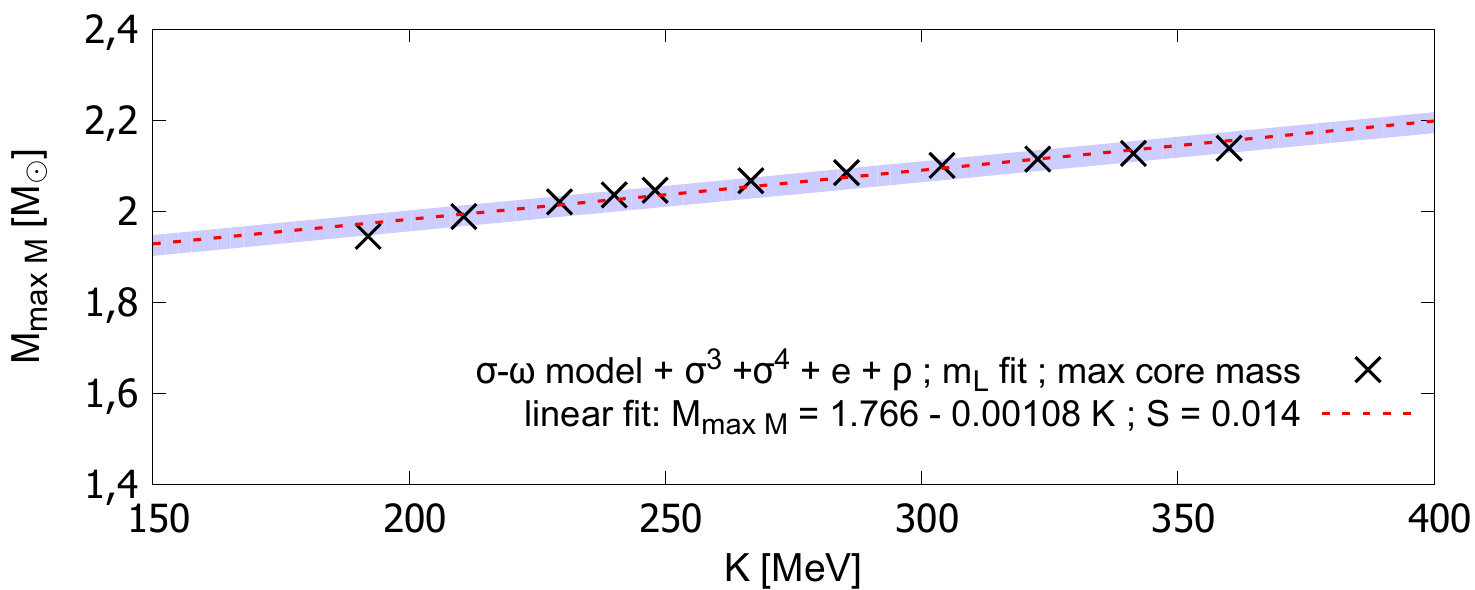}
\caption{}
\label{egy}
\end{subfigure}
\begin{subfigure}{0.5 \textwidth}
\centering
\includegraphics[width=0.95\linewidth]{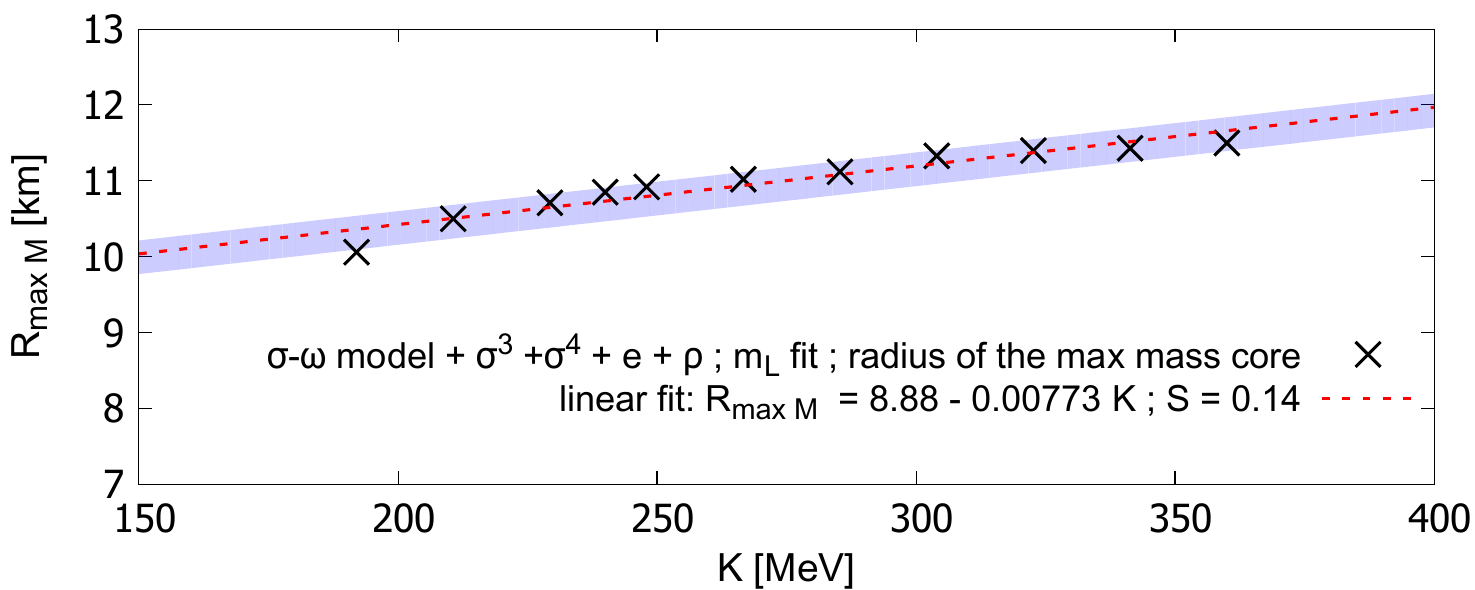}
\caption{}
\end{subfigure}
\begin{subfigure}{0.5 \textwidth}
\centering
\includegraphics[width=0.95\linewidth]{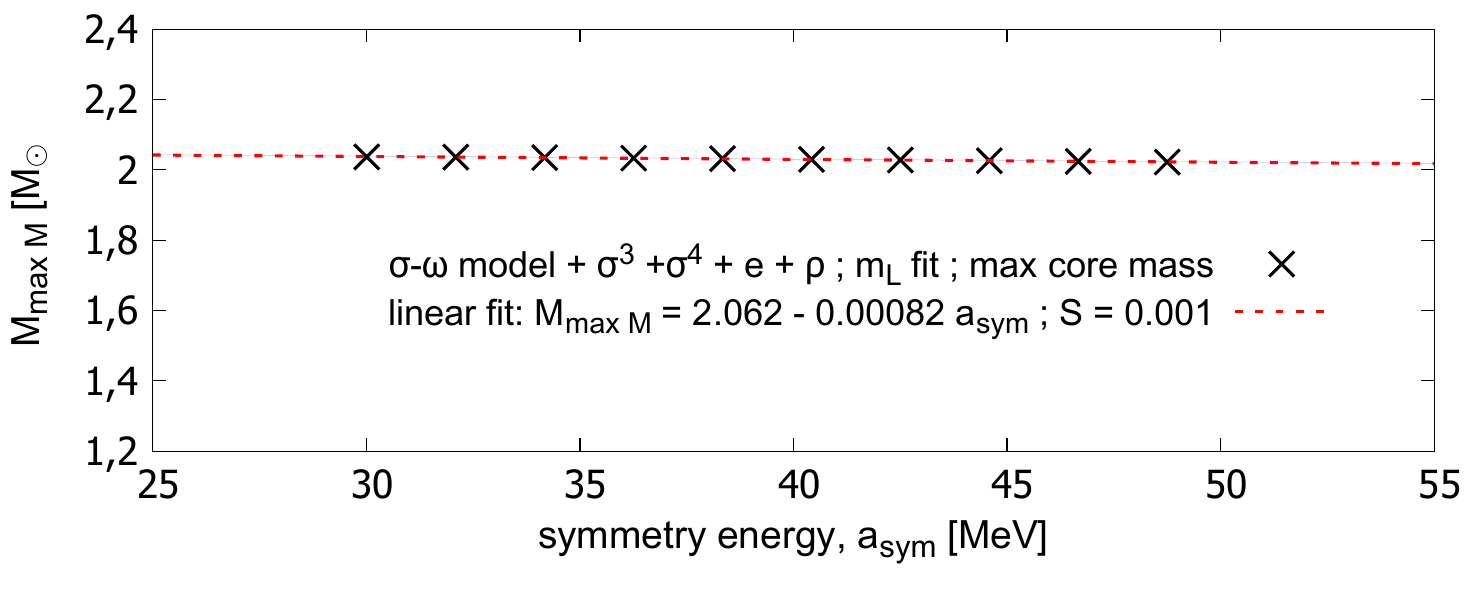}
\caption{}
\label{egy2}
\end{subfigure}
\begin{subfigure}{0.5 \textwidth}
\centering
\includegraphics[width=0.95\linewidth]{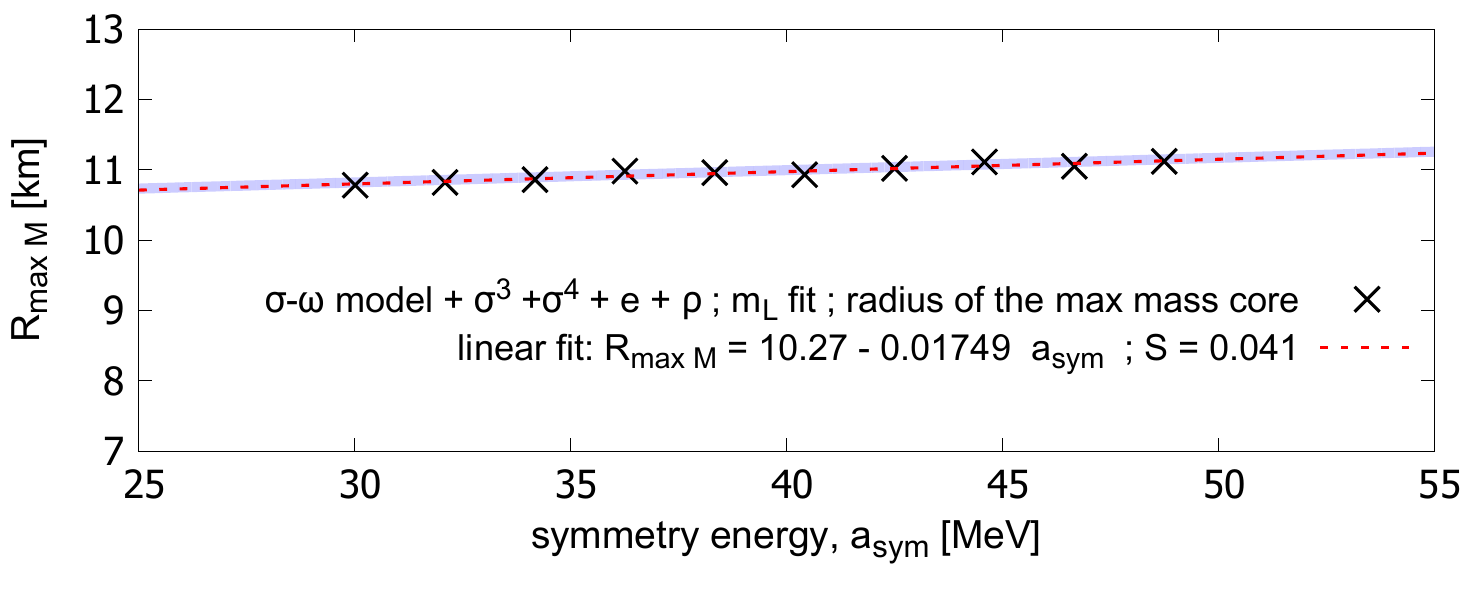}
\caption{}
\end{subfigure}
\caption{Maximum mass star parameters as functions of different nuclear parameters. The data is drawn as black crosses, the best linear fit is plotted as a dashed red line and finally the band of error is drawn as a blue region around the red line. Observation data points are explained in the text.}
\label{fig:mm_fits}
\end{figure*}

The results of these calculations for the maximal mass star can be seen on Figs.~\ref{fig:mm_fits}.
Interestingly, linear approximation works very well to describe how the mass and radius of the MMS 
depend on the value of Landau mass of nucleons, on the compression modulus, and on the value of the symmetry energy. The best linear fit is plotted as a red dashed line and a blue band of error is also drawn which in some cases is hardly noticeable. The linear approximation works perfectly for MMS where the error of the linear fit is the smallest. The linear relation is used to translate the mass observations to predictions of the values of nuclear parameters. An errorbar in $M$, results in uncertainty in the $m_L$ values. In case of the $R(m_L)$ function this works reversely: the so-determined uncertainty in $m_L$ can be transferred to the errorbars in the radius. We also indicated the masses of known MMS objects on Fig.~\ref{fig:mm_fits}, with the corresponding observational and theoretical uncertainties. See details in Section~\ref{sec:dis}.

Comparing the parameters of the fitted linear functions sheds light on how strongly different nuclear parameters influence the neutron star parameters. The fitting parameters can be found in Table~\ref{tab:lin_fit} for MMS.

To make the comparison more obvious in Table~\ref{tab:lin_fit} every nuclear parameter is measured in its respective saturation value. The connections between the fits in Figs.~\ref{fig:mm_fits} together with the data in Table~\ref{tab:lin_fit} is the following: 
%
\begin{equation}
\begin{aligned}
f(\xi)& = (a' \xi_{0}) \left( \frac{\xi}{\xi_{0}} \right) + b  \ , \\
a & = a' \xi_{0} ,
\end{aligned}
\label{eq:fit_eq}
\end{equation}
%
where parameter, $a'$ is the slope of the linear function plotted on Fig.~\ref{fig:mm_fits}, 
while the rescaled slope, $a$ used in the tables and $\xi_{0}$ is the saturation value of the tuned nuclear parameter. We note specific equations are explicitly given in Figs.~\ref{fig:mm_fits}. 
%
%
%

%
%
%
%
%
%
%

Applying this re-scaling for MMS values, the slopes of the linear functions correspond to the same percentage of change in the different parameters. Larger slope means larger variation in neutron star mass or radius for the same relative deviation in the studied nuclear parameters. 

It can be concluded that in case of the MMS the mass and the radius are much more influenced by the nuclear effective mass than the nuclear compressibility and varying the symmetry energy causes even smaller changes. However as it can be seen on Fig.~\ref{fig:mr} that including the $\beta$-equilibrium (and hence symmetry energy) in the calculations is important because the interaction terms introduced into the symmetric Lagrangian cause a change in the maximum star mass and radius. An important consequence of the above mentioned properties is that the precise value of the symmetry energy is not relevant in the extended $\sigma$-$\omega$ model to study the $M$-$R$ diagram because varying its value barely changes the $M$-$R$ curve.

\begin{table}[h]
\caption{\label{tab:lin_fit} Parameters of the linear fits describing  how observable parameters of neutron stars depend on various parameters of nuclear matter. A neutron stars considered from the $M$-$R$ diagram is the maximum mass star and 'SD' is the standard deviation of the fit.}
\begin{center}
\begin{tabular}{@{}rccc}
\hline \hline
Fit parameters for MMS         & a         & b        & SD     \\    
\hline
          & \multicolumn{3}{c}{\textbf{Effective mass}}                           \\
\textbf{Mass dependence}   & -3.38     & 5.4      & 0.008   \\  
\textbf{Radius dependence} & -8.1    & 19.04   & 0.17      \\   
\hline
         & \multicolumn{3}{c}{\textbf{Compression modulus}}                      \\
\textbf{Mass dependence}   & 0.26     & 1.77      & 0.014    \\ 
\textbf{Radius dependence} & 1.86      & 8.88      & 0.14     \\    
\hline
         & \multicolumn{3}{c}{\textbf{Symmetry energy}}                         \\
\textbf{Mass dependence}   & 0.03      & 2.062     & 0.001   \\ 
\textbf{Radius dependence} & 0.57     & 10.27      & 0.041  \\   
\hline \hline
\end{tabular}
\end{center}
\end{table}

Table~\ref{tab:lin_fit} also reveals that linear approximation works better to describe the mass dependence of the MMS than to describe radius dependence since the standard errors of the fits were the smallest in these cases. 
The mass of the MMS depends approximately 10 times more strongly on nucleon Landau mass than on the compression modulus and it depends $\sim$100 times more strongly on Landau mass than on the value of symmetry energy. In the case of the radius of MMS it depends $\sim$5 times more on the Landau mass then on the compression modulus and $\sim$10 times more on Landau mass then on the symmetry energy. It can be stated generally that the compression modulus and symmetry energy influences the radius of MMS stronger than their mass, because the slopes in the table has the highest relative values. 

We note that this has an interesting application if one takes into consideration that measurements regarding neutron star masses has higher precision than radius measurements. At a given accuracy this suggest that the  mass constraints provided by measurements can be met with tuning only one parameter, the Landau mass, in these models.  

To summarize: different microscopic nuclear parameters influence different parts of the $M$-$R$ diagram differently in the case of the extended $\sigma$-$\omega$ model. The value of the Landau effective mass influences neutron star parameters much stronger than the other two parameters. This means that high precision radius and mass measurements seems to be crucial for the fitting of the models, because these would make it possible to separate the smaller effect of the compression modulus and symmetry energy from the robust effect of Landau mass. 

\section{The effect of the low-energy EoS on the $M$-$R$ diagram}

In our calculations the effect of the low-energy, neutron stars crust EoS is not taken into account to focus on the effects of the high-density nuclear EoS. The inclusion of the crust EoS would introduce several new parameters which would make it challenging to extract the information regarding the high-density nuclear matter from the results. Our calculations are conservative regarding the mass and the radius of the neutron stars. To get a better picture how our result can be used in more realistic scenario for example in neutron star observations we calculated the $M$-$R$ diagram of the best fitting model in three ways. 
\begin{itemize}

\item[(i)] Conservative: the TOV equations are integrated using the high-density nuclear EoS. In this case the model is the extended $\sigma$-$\omega$ model with both the 3\textsuperscript{rd} and 4\textsuperscript{th} order interaction terms are fitted on the value of the Landau mass. 

\item[(ii)] Core calculation: the integration of the TOV equations is stopped at the edge of the core and the crust in not calculated. This is the above mentioned core approximation. 

\item[(iii)] Core with crust calculation: the BPS equations of state is used to describe the crust and finish the integration of the TOV equations at low densities~\cite{BPS}. 
\end{itemize}

\begin{figure}[!h]
\centering
\includegraphics[width=0.45\textwidth]{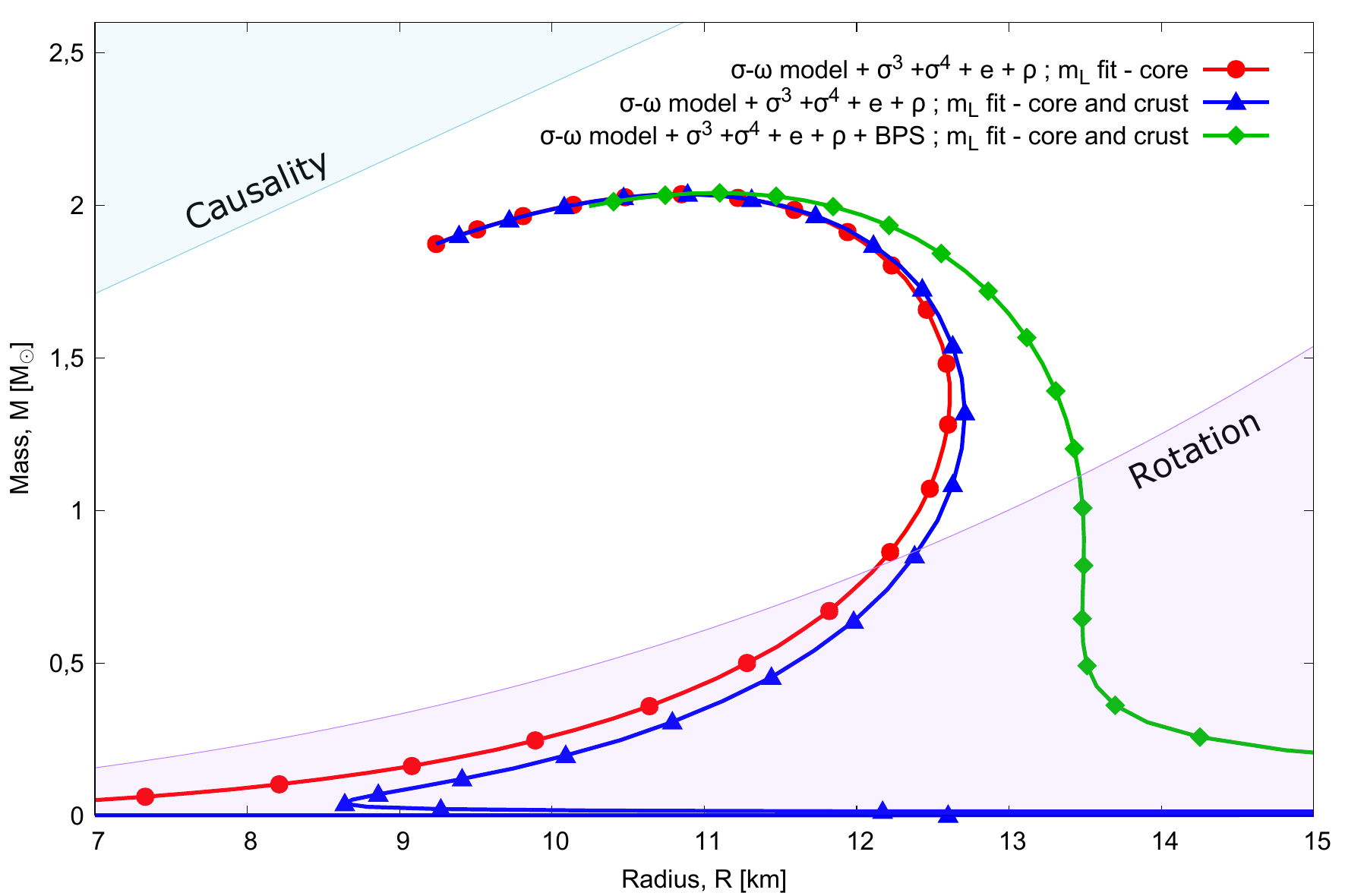}
\caption{\label{fig:crust_comp}
Three different calculations of the $M$-$R$ diagram corresponding to extended $\sigma$-$\omega$ model fitted on the Landau mass. The red line (full circle) is the core approximation, the most conservative calculation (i). The blue line (triangle) corresponds, to calculation (ii), where only the high energy nuclear equations is used during the integration of the TOV equations. The green line (diamonds) corresponds to the calculation (iii), where the $\sigma$-$\omega$ model is used to describe the core of the stars and in the crust it is switched to the BPS low-energy EoS by~\cite{BPS}. 
}
\end{figure}

The comparison of  these three ways of calculating the $M$-$R$ diagram is shown on Figure~\ref{fig:crust_comp}. 
As it can be seen the crust EoS influences the $M$-$R$ diagram in the region of low- and intermediate-mass neutron stars more than the high-mass neutron stars in accordance with other results summarized by~\cite{lotmodell}. The inclusion of a more realistic crust EoS increased the maximum radius of the neutron stars considerably but it has small effect on the value of the maximum star mass. This means that our results regarding the maximum mass stars can carry on for more realistic cases and neutron star measurements. From the microscopic parameters of nuclear matter we consider here the nucleon effective mass has much larger influence on the maximum star mass than the others. The calculated curves presented on Figure~\ref{fig:crust_comp} suggest that this result is more general as the crust EoS has little influence on the value of the maximum star mass. This means that observed neutron star parameter measurements can be used to set the correct value of the effective nucleon mass in the $\sigma$-$\omega$ model because the maximum neutron star mass is less sensitive to the microscopic parameters beyond the effective mass. 
%
%

As mentioned above the effect of other two microscopic parameters studied here can be seen best on the radius of the star. Since the effects of the low energy EoS are the largest in the same region the study of the effect of compression modulus and symmetry energy in realistic neutron star data is challenging and has to be carried out together with the study of the effect of the low-energy EoS. These results can be understood intuitively since both the compression modulus and symmetry energy characterize the nuclear EoS at low energies just as the EoS describing the neutron star crust according to~\cite{lotmodell}. 

\section{Discussion: application of the results in model fitting}
\label{sec:dis}

All constraints available for the nuclear matter model from compact star observations are suffer from the {\it masquarade} problem, thus investigation of specific models and parameter values are uncertain. Since the accuracy of astronomical instruments developed well, the amount of data has been increased and further rich data are expected soon, providing quantitative uncertainties and clear parameter dependencies in the era of the multi-channel astronomy. However, the first pulsar, PSR J0030+0451, with joint mass and radius measurement has been added to the $M$-$R$ diagram~\cite{niser:2019}, we are still far from the high-statistics to provide strong statement on the nuclear matter EoS.

From all of the nuclear parameters considered here, effective mass stands out in the sense that it is connected to two independently measurable physical parameters of symmetric nuclear matter: effective mass and Landau mass. This means that in relativistic mean field models the nucleon effective and Landau mass are not independent, and therefore can not be fitted simultaneously. As stated above models which are parametrized to reproduce the measured value of nucleon Landau mass produce neutron stars with mass and radius which is the closest to the astrophysical observational data. It was also shown that the maximum neutron star mass depends approximately linearly on nucleon Landau mass. Using this connection the model can be fitted very simply to reproduce measured data regarding maximal-mass as follows. 

Based on the existing data and our assumptions the highest mass of a neutron star is a well-determined and best candidate to use for experimental validation of our results. For this reason, we use the highest-mass pulsar data for the verification, with small errors due to the precision of the analysis method: PSR J1614$-$2230~(\cite{PSRJ16}), PSR J0348+0432~(\cite{PSRJ03}), and PSR J0740+6620~(\cite{PSRJ07}). Based on the data in Table.~\ref{tab:maxM-lm} and fits in Fig.~\ref{fig:mm_fits}, the connection between Landau mass in MeV and maximum star mass in M$_{\odot}$ and the radius of the maximal-mass neutron star in km units is the following,
\begin{eqnarray}
M_{maxM}[\textrm{M}_{\odot}] = 5.418 - 0.0043 \, m_L [\textrm{MeV}] \ , \label{eq:maxmass_M-lm} \\ 
R_{maxM}[\textrm{km}] = 19.04 - 0.0104 \, m_L [\textrm{MeV}]   \ . \label{eq:maxmass_R-lm} 
\end{eqnarray}
The fits have theoretical uncertainties which was characterized by the $\chi^2$-method for $M_{maxM}[\textrm{M}_{\odot}]$ and $R_{maxM}[\textrm{km}]$ these are $0.8\%$ and $17\%$, respectively. 

Supported by the Refs.~\cite{PSRJ17,LucianoMaxcikk,Ozel:2016oaf} one can assume, PSR J1614-2230, PSR J0348+0432, and PSR J0740+6620 masses are (close) to maximal mass pulsars, in which case Landau mass can be calculated from measured $M_{maxM}$ values using eq.~\eqref{eq:maxmass_M-lm}, then $R_{maxM}$ can be also obtained from eq.~\eqref{eq:maxmass_R-lm}. The main motivation for this is that the maximal mass is unsensitive to the model variants as on Fig.~\ref{fig:crust_comp}, especially after fixing the relevant physical parameter, $m_L$ and deal with the proper bunch on Fig.~\ref{fig:mr}. For this reason the core of the neutron star will dominate and determine the overall nuclear matter EoS~\cite{Demorest}.

Using this assumption we calculated the Landau mass value based on the mass of the pulsars in each case using eq.~\eqref{eq:maxmass_M-lm}. Then we inserted the Landau mass values obtained from the pulsar data into eq.~\eqref{eq:maxmass_R-lm} to calculate the corresponding radius values for the maximum mass stars. Results including both observational and calculated data and theoretical uncertainties are listed in Table~\ref{tab:maxM-lm} below, where measured values are denoted by '$^{\ast}$'. Landau mass values including their errors are plotted on Fig.~\ref{sfig:mm_mL_m}. Vertical errors are from the measured data, while horizontal errors includes theoretical uncertainties together with the error of the observation as well. 
It is interesting to see, that predictions are overlapping within the errors, and provide an average value, 
\begin{equation}
m_{L} = 786.75 \pm{7.77}~\textrm{MeV},
\label{eq:landau_mass_res}
\end{equation} 
confirming the assumption on the relevance of the maximal mass neutron star fit. 

The predicted values of the maximum mass neutron star radii and their uncertainties are plotted on Fig.~\ref{sfig:mm_mL_r}. All three values predict neutron star radius between $10.5 - 12 \, \text{km}$, highlighting the {\sl masquarade} problem in this one-parameter system: the value of the macroscopic parameters (here the radii of MMS) is very similar for  different values of the microscopic parameter (here the Landau mass). Here this property can be used to give a robust estimation for the radius of the MMS considering the fact, that as it can be seen on Fig.~\ref{fig:crust_comp} in the case of MMS other factors like the crust has minimal effect and the effect of the core is dominated by the value of the Landau mass. The magnitude of uncertainty corresponding to the observation based value of microscopical nuclear parameters, c.f. {\sl masquarade} problem, including all the theoretical uncertainties and errors, is about $10\%$, which corresponds to our recent estimates in~\cite{symwal,Posfay:2017cor}.  

%
\begin{table*}[h!]
\caption{\label{tab:maxM-lm} The Landau mass value calculated via eq.~\eqref{eq:maxmass_M-lm} from measured pulsar mass data, assuming that these are maximal-mass neutron stars. Radius estimates of the maximal-mass neutron stars are also calculated from eq.~\eqref{eq:maxmass_R-lm}. Error estimates correspond to one standard deviation. }
\begin{center}
\begin{tabular}{lccc}
\hline \hline 
\textbf{Pulsar}   & $M_{maxM}$[M$_{\odot}$] & $R_{maxM}$[km] & $m_L$[MeV]         \\
\hline
PSR J0740+6620 & 2.08$ \pm{0.07} \ast $ & 11.06$\pm{0.79}$   & 769.89$\pm{17.96}$  \\
PSR J0348+0432 & 2.01$\pm{0.04} \ast$ & 10.89$\pm{0.79}$   & 786.04$\pm{12.18}$  \\
PSR J1614$-$2230 & 1.97$\pm{0.04} \ast$ & 10.79$\pm{0.80}$   & 795.26$\pm{12.21}$  \\
\hline \hline
\end{tabular}
\end{center}
\end{table*}

\begin{figure}[!h]
\begin{center}
\includegraphics[width=0.49\textwidth]{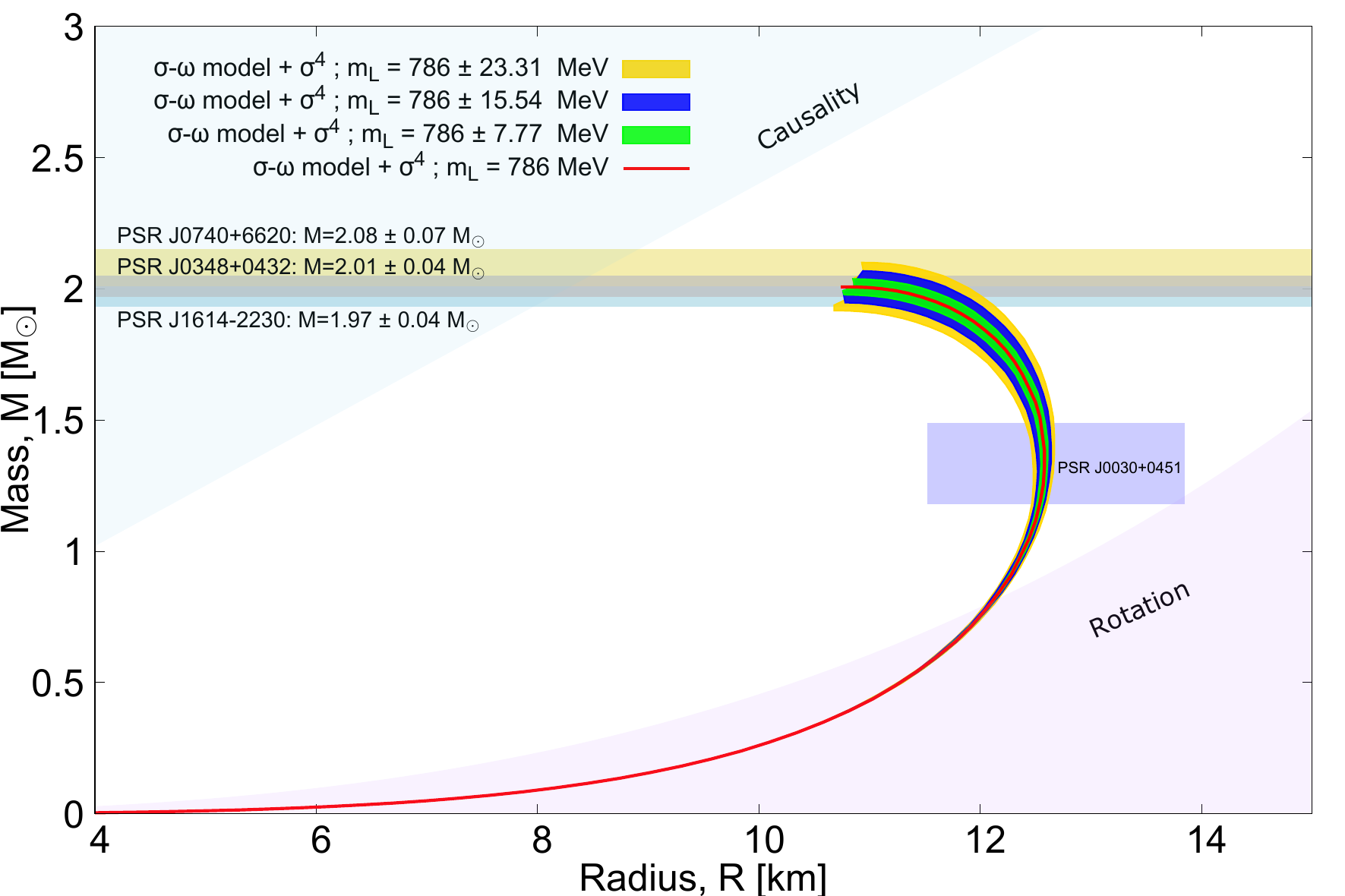}
\end{center}
\caption{\label{fig:ml_fit} The filled area between the curves corresponds to the error in Landau mass in \eqref{eq:landau_mass_res} which is derived from the error of the mass measurement of maximal mass pulsars: PSR J1614$-$2230, PSR J0348+0432, and PSR J0740+6620.}
\end{figure}

To demonstrate how good are the linear approximations in equation~\eqref{eq:maxmass_M-lm} and in relation with this in equation~\eqref{eq:maxmass_R-lm} the model was fitted two times: for the maximum possible value of the Landau mass and for the lowest one. The different EoS corresponding to these fits than used to solve the TOV equations and calculate the $M$-$R$ diagrams for each fit. The results can be seen on Fig.~\ref{fig:ml_fit}.
The deviance in the $M$-$R$ diagram caused by the deviance in Landau mass is the largest for stars  with mass above $1.5 \, M_{\odot}$ and it is zero below $1 \, M_{\odot}$. The observation data regarding the maximum star mass is reproduced up to two digit precision using the linear approximation in equations~\eqref{eq:maxmass_M-lm} and~\eqref{eq:maxmass_R-lm}.  
%
 
It is interesting to see, although the {\sl masquarade problem}, the mean microscopical nuclear values are pretty precise and physically relevant on the other hand together with the macroscopical observables provide a comprehensive picture with a strong predictive power. This governed us to the $R_{maxM}$ radii predictions of maximal mass pulsars: PSR J1614$-$2230, PSR J0348+0432, and PSR J0740+6620 as listed in Table~\ref{tab:maxM-lm} above. As it was proposed by~\cite{Miller:2016}, NICER collecting data on PSR J1614$-$2230, thus our prediction hopefully can be cross checked soon with~\cite{Wolff:2019}.
 
\section{Summary}

We investigated the {\sl masquarade problem}: the effect of nuclear model parameters on the compact star's macroscopic observable parameters, mass and radius were analyzed in details. We studied the extended $\sigma$-$\omega$ model Lagrangian with three different interaction terms and optimized nuclear parameter values to calculate the dense nuclear matter equation of state. These were inserted to a static, spherically symmetric space-time, and using the Tolman\,--\,Oppenheimer\,--\,Volkoff equation, observable parameters of neutron stars were calculated.

We investigated, that both the 3\textsuperscript{rd} and 4\textsuperscript{th} order scalar terms are required for the best description of the nuclear matter equation of state. In this case the effective mass can be chosen optimally in accordance with the nuclear experimental data. We identified, that the optimal Landau effective mass is the most relevant physical parameter modifying the macroscopic observable values, followed by the compressibility and symmetry energy terms. The variation of these nuclear parameters generate one-one order of magnitude smaller effect on the maximal mass value of the MMS, respectively,
$$ \Delta M_{max}(\delta m_L) \overset{10 \times}{ >} \Delta M_{max}(\delta K) \overset{10 \times}{ >}  \Delta M_{max}(\delta a_{sym}) , $$
Moreover, we obtained that in case of the maximum mass compact star the mass and radius of the star's core depend linearly well on the studied microscopic nuclear parameter values within the physically relevant parameter range. We also estimated the magnitude of the maximal mass pulsars' radii of PSR J1614$-$2230, PSR J0348+0432, and PSR J0740+6620 with the theoretical uncertainties arising from the interaction terms of the nuclear matter and their parameter values choice. 

Using our results regarding the {\sl masquarade} problem we estimated the value of the Landau effective mass based on modern pulsar mass measurements. The result is $m_{L} = 786.75\pm{7.77}$~MeV which is within agreement with nuclear physics data. The interplay between different nuclear parameters is not taken into account in this calculation apart from the obtained effects' orders. Other studies where the effect of additional parameters like for example the compression modulus is taken into account yield a similar result within $1.5\sigma$ ~\cite{Szigeti} and $4.5 \sigma$~\cite{Szigeti_unpublished} respectively.
For further investigation of this assumption we found similar results by a Bayesian analysis applying different experimental and observational data constraints in~\cite{Alvarez}. Based on this preliminary results we can give magnitude-estimation of the radius of the maximum mass star in a more general way -- without restricting our results on MMS. 


\begin{acknowledgements}
Authors acknowledge the remarks and fruitful discussions for David Blaschke, P\'eter Kov\'acs, Andr\'as Patk\'os, and Zsolt Szép. This work is supported by the Hungarian Research Fund NKFIH (OTKA) under contracts No. K120660, K123815, K135515, and COST actions PHAROS (CA16214) and THOR (CA15213). Authors also acknowledges the computational resources for the Wigner Scientific Computing Laboratory (WSCLAB) the former Wigner GPU Laboratory.
\end{acknowledgements}

\bibliographystyle{apj}

\bibliography{walref07}

\end{document}